\documentclass[11pt]{article}
\usepackage[utf8]{inputenc}
\usepackage{amsmath, amssymb}
\usepackage{graphicx}
\usepackage{authblk}
\usepackage{geometry}
\usepackage{fancyhdr}
\usepackage{newunicodechar} 

\title{Capsule-ConvKAN: A Hybrid Neural Approach to Medical Image Classification\thanks{Preprint version. Accepted to IEEE SMC 2025.}}

\author[1]{Laura Pituková}
\author[1]{Peter Sinčák}
\author[2]{László József Kovács}
\author[3]{Peng Wang}

\affil[1]{Department of Cybernetics and Artificial Intelligence, Technical University of Košice}
\affil[2]{Institute of Information Science, University of Miskolc}
\affil[3]{Department of Computer Science and Engineering, University of Connecticut}

\date{}  

\begin{document}

\maketitle
\thispagestyle{fancy} 
\thispagestyle{empty}
\pagestyle{empty}

\begin{abstract}
  This study conducts a comprehensive comparison of four neural network architectures: Convolutional Neural Network, Capsule Network, Convolutional Kolmogorov–Arnold Network, and the newly proposed Capsule-Convolutional Kolmogorov–Arnold Network. The proposed Capsule-ConvKAN architecture combines the dynamic routing and spatial hierarchy capabilities of Capsule Network with the flexible and interpretable function approximation of Convolutional Kolmogorov–Arnold Networks. This novel hybrid model was developed to improve feature representation and classification accuracy, particularly in challenging real-world biomedical image data. The architectures were evaluated on a histopathological image dataset, where Capsule-ConvKAN achieved the highest classification performance with an accuracy of 91.21\%. The results demonstrate the potential of the newly introduced Capsule-ConvKAN in capturing spatial patterns, managing complex features, and addressing the limitations of traditional convolutional models in medical image classification.

\end{abstract}

\section{INTRODUCTION}

Rapid advances in deep learning have spurred the creation of novel neural‑network architectures for image recognition, feature extraction and pattern analysis. The Kolmogorov‑Arnold Network (KAN) has attracted interest for its strong theoretical foundation in function approximation, whereas the Convolutional Neural Network (CNN) has long dominated visual tasks thanks to its ability to capture spatial hierarchies. Merging these ideas led to the Convolutional Kolmogorov‑Arnold Network (ConvKAN), which couples convolutional feature extractors with KAN‐style spline activations 

Capsules add an additional layer of structure: the Capsule Network (CapsNet) introduces dynamic routing and explicit pose vectors, enabling the model to preserve part–whole relationships and handle viewpoint changes. Integrating capsules into ConvKAN yields the Capsule‑Convolutional Kolmogorov‑Arnold Network (Capsule‑ConvKAN)—a hybrid that, to our knowledge, has not yet been systematically studied. This work closes that gap by comparing ConvKAN and Capsule‑ConvKAN in terms of performance, efficiency and robustness on image‑analysis tasks.

The Kolmogorov–Arnold Representation Theorem (Section \ref{subsec:k-a_rt}) underpins KAN’s smooth, interpretable function blocks, while convolutional layers remain the work‑horse for spatial feature extraction. By fusing CNN and KAN components, ConvKAN inherits both expressive power and interpretability; the addition of capsule mechanisms further augments the model’s ability to capture hierarchical spatial information. We therefore investigate whether Capsule‑ConvKAN can outperform its purely convolutional and spline‑based predecessors, especially on challenging, real‑world biomedical imagery.

The foundational work by Liu et al. \cite{liu2024kan}, introduced KANs as a novel neural architecture that replaces conventional weights with learnable spline functions. This approach enables greater flexibility and interpretability, while also improving accuracy and reducing parameter count compared to traditional multilayer perceptrons.

Building on this concept, Drokin \cite{drokin2024kolmogorovarnoldconvolutionsdesignprinciples}, presents design principles and empirical studies on Kolmogorov-Arnold convolutions, further advancing the understanding of this hybrid approach in computer vision tasks. The author proposed the ConvKAN, incorporating KAN layers into CNNs. Their architecture demonstrated enhanced robustness, interpretability, and classification performance on standard datasets such as MNIST, CIFAR-10/100, Tiny ImageNet, ImageNet1k, and HAM10000.

Roy et al. \cite{10903072} applied ConvKAN to the domain of medical imaging, specifically for brain tumor classification using MRI scans. Their study integrated KAN with convolutional feature extraction and demonstrated improved accuracy and reliability in tumor detection. The results highlight the applicability of ConvKAN in high-stakes tasks such as medical diagnosis.

While Multilayer Perceptron (MLP) build upon the Universal Approximation Theorem \cite{lu2020universal},
\begin{equation}
    f(x) \approx \sum_{i=1}^{N} \alpha_i \sigma(\mathbf{w}_i \cdot \mathbf{x} + b_i)
\end{equation}

which states that a neural network with at least one hidden layer, a sufficient number of neurons, and a continuous activation function, can approximate any continuous function to any desired level of accuracy, provided the weights are properly adjusted, KAN builds upon the principles of the Kolmogorov-Arnold Representation Theorem, detailed in Section \ref{subsec:k-a_rt}, which enables the decomposition of multivariate functions into sums of univariate functions \cite{liu2024kan}. 
\begin{figure}[h!]
\centering
\includegraphics[width=0.5\textwidth]{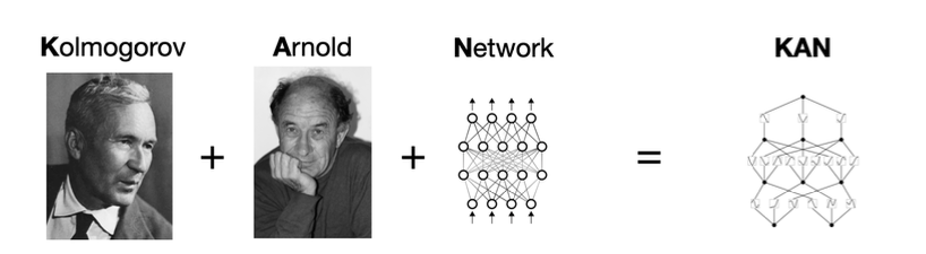}  
\caption{Kolmogorov-Arnold Network \cite{liu2024kan}}
\label{fig:kanrt}
\end{figure}

This theory guides the design of KAN, allowing it to model complex relationships while retaining an interpretable structure:
\begin{equation}
    KAN(x) = (\phi_3 \cdot \phi_2 \cdot \phi_1)(x)
\end{equation}

while MLP is typically represented as a series of transformations that apply weights and activation functions at each layer:
\begin{equation}
MLP(x) = (W_3 \cdot f_2 \cdot W_2 \cdot f_1 \cdot W_1)(x)
\end{equation}

KAN is interpretable due to its unique design, which contrasts with traditional neural networks. Unlike traditional neural networks, KAN layers use B-splines as learnable activation functions \cite{genet2024tkan}, introducing a new paradigm in function approximation.

\subsection{Theoretical Foundations of KAN}\label{subsec:k-a_rt}
Kolmogorov-Arnold Representation Theorem says, that any multivariate continuous function "f" defined on a bounded domain can be represented as a finite composition of continuous functions of a single variable \cite{schmidt2021kolmogorov}, combined using the binary operation of addition.

\begin{equation}
f(\mathbf{x}) = f(x_1,\cdot\cdot\cdot,x_n) = \sum_{q=1}^{2n+1}\phi_q\left(\sum_{p=1}^{n} \phi_{q,p}(x_p)\right)
\end{equation}

In machine learning it means, instead of writing the equation as “multi-variate” function:
\begin{equation}
Result = F(x, y, z)
\end{equation}
where the result depends directly on the joint relationship between all the features 
\(x\), \(y\), \(z\), we can represent it as an aggregate of multiple univariate functions. Each univariate function independently depends on a single feature, allowing us to express the result as:

\begin{equation}
Result = f(\phi_x(x) + \phi_y(y) + \phi_z(z))
\end{equation}
where  \(\phi_x(x)\), \(\phi_y(y)\), and \(\phi_z(z)\) represent the transformations of each feature \cite{somvanshi2024survey}. This decomposition enables a more modular and interpretable modeling approach, even enhancing both efficiency and clarity during the learning process. 

The KAN layer applies univariate transformations independently to each input feature \cite{liu2024kan}, combines their contributions through sum, and generates the final prediction.

\subsection{KAN Layer} \label{subsec:layer}
In a KAN layer, the input features are processed by these B-splines independently. The transformation process can be represented as a matrix operation, where the matrix elements are the B-spline functions applied to the input features \cite{liu2024kan2}.

Given \( n_{\text{in}} \) input features \( x_1, x_2, \dots, x_{n_{\text{in}}} \), KAN layer can be represented as:

\begin{equation}
    y = \sum_{p=1}^{n_{\text{in}}} \Phi_p(x_p)
\end{equation}

where \( \Phi_p(x_p) \) represents the transformation of the \( p \)-th input feature \( x_p \), and the sum of these transformed features yield the final output \( y \).

\
\begin{equation}
\begin{bmatrix}
\Phi_{11} & \Phi_{12} & \dots & \Phi_{1n} \\
\Phi_{21} & \Phi_{22} & \dots & \Phi_{2n} \\
\vdots & \vdots & \ddots & \vdots \\
\Phi_{m1} & \Phi_{m2} & \dots & \Phi_{mn}
\end{bmatrix}
\begin{bmatrix}
x_1 \\
x_2 \\
\vdots \\
x_n
\end{bmatrix}
\
\end{equation}
Summation of the matrices:
\begin{equation}
\begin{bmatrix}
\sum \Phi_{1q}(x_q) \\ \sum \Phi_{2q}(x_q) \\ \vdots \\ \sum \Phi_{mq}(x_q) \\
\end{bmatrix}
\end{equation}
As illustrated in Figure \ref{fig:layers}, KAN layers' input features are transformed through a series of univariate functions to produce the final output:
\begin{figure}[h!]
\centering
\includegraphics[width=0.48\textwidth]{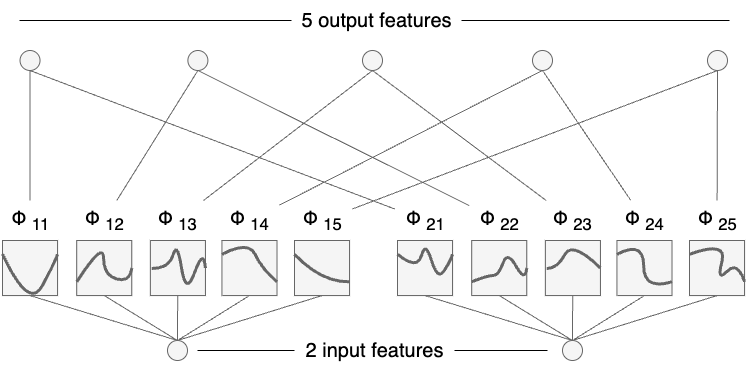}  
\caption{Visual representation of KAN layers}
\label{fig:layers}
\end{figure}
\\
The univariate functions \(\phi(x)\) have trainable parameters. These parameters are adjusted during training \cite{somvanshi2024survey}, allowing the KAN to model complex relationships between inputs.
Kan's architecture can be stacked in multiple layers for deeper representations. Each additional layer can represent more complex relationships between features.

\subsection{Basis spline (B-spline)} \label{subsec:spline}
The KAN layer enhances non‑linear modeling by replacing fixed activations—such as ReLU ($\max(0,x)$) or SiLU ($x\cdot\text{sigmoid}(x)$)—with learnable B‑spline activations whose shapes are optimized directly from the data, allowing the network to transform feature maps more flexibly and to discriminate complex patterns more effectively \cite{liu2024kan}.

\begin{figure}[h!]
\centering
\includegraphics[width=0.45\textwidth]{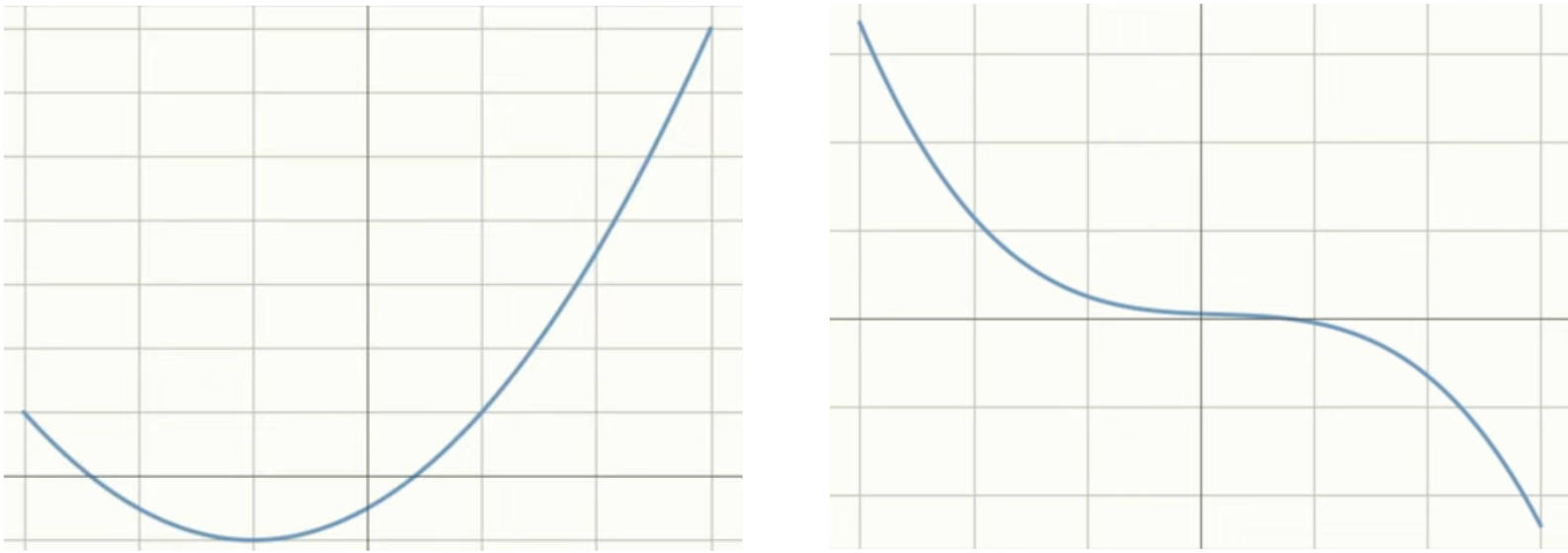}  
\caption{Example of splines}
\label{fig:splines}
\end{figure}

As illustrated in Fig. \ref{fig:splines}, a B‑spline is a piecewise‑defined, smooth curve expressed as a linear combination of basis functions determined by a set of control points \cite{vaca2024kolmogorov}.  Thanks to their locality—changes to one control point affect only a small neighbourhood—B‑splines provide fine‑grained, task‑specific non‑linear transformations while preserving interpretability; this local control also yields highly accurate approximations in low‑dimensional settings and enables the construction of intricate function shapes without the artifacts typical of fixed activations \cite{cheon2024demonstrating}.

The construction of B-splines can be differentiable, which is very important for backpropagation. Differentiability enables gradient-based optimization methods \cite{moradi2024kolmogorov}, allowing B-splines to effectively contribute to model learning.

B-splines can be defined mathematically:
\begin{itemize}
    \item The B-spline basis function, as \(N_{i,0}(t)\) is 1 if the parameter \(t\) lies between \(t_i\) and \(t_{i+1}\), and 0 otherwise.
    \begin{equation}
    N_{i,0}(t) = 
    \begin{cases} 
    1 & \text{if } t_i \leq t < t_{i+1} \text{ and } t \leq t_{i+1} \\
    0 & \text{otherwise}
    \end{cases}
    \end{equation}  
\end{itemize}
\begin{itemize}
    \item Defining the higher-order B-spline basis functions as \( N_{i,j}(t) \). For \( j > 0 \), it is a weighted combination of two lower-order basis functions \cite{liu2024kan}, \( N_{i,j-1}(t) \) and \( N_{i+1,j-1}(t) \), with the weights being fractions of the distance between the knot points.
\end{itemize}
\begin{equation}
N_{i,j}(t) = \frac{t - t_i}{t_{i+j} - t_i} N_{i,j-1}(t) + \frac{t_{i+j+1} - t}{t_{i+j+1} - t_{i+1}} N_{i+1,j-1}(t), 
\end{equation}
where \( j = 1, 2, \dots, p \). Then the curve defined by:
\begin{itemize}
    \item The curve \(C(t)\) is defined as a weighted sum of control points \(P_i\) and the B-spline basis functions \(N_{i,p}(t)\):
\end{itemize}
\begin{equation}
C(t) = \sum_{i=0}^{n} P_i N_{i,p}(t)
\end{equation}
where \( i = 0, 1, \dots, n \), and \( N_{i,p}(t) \) is a B-spline basis function of order \(p\).

For training and evaluation, we used PyTorch. The training process involved batch gradient descent, where we optimized the model using AdamW with a learning rate scheduler to dynamically adjust learning rates based on validation performance.

During each epoch, we monitored training loss, validation loss, and validation accuracy. In the architecture was applied a dropout layer with a dropout rate of 30\%, to prevent overfitting. The training employed an early stopping strategy with a patience of 5 epochs, meaning training stopped if the validation loss did not improve for 5 consecutive epochs. Additionally, we applied gradient clipping to stabilize training and avoid exploding gradients. 

\subsection{CNN}
\begin{figure}[h!]
\centering
\includegraphics[width=0.48\textwidth]{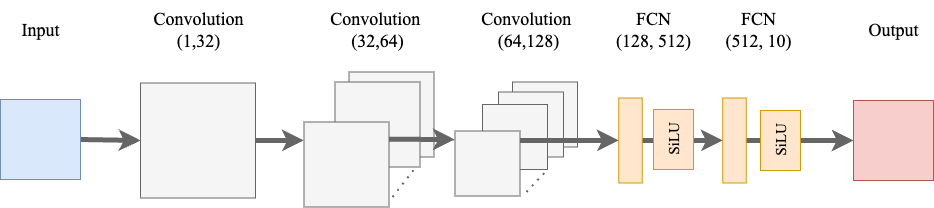}  
\caption{Architecture of Convolutional Neural Network}
\label{fig:cnn}
\end{figure}

CNN architecture, as illustrated in Fig. \ref{fig:cnn}, consists of a feature extractor followed by fully connected layers for classification. The feature extractor is composed of three convolutional layers, each followed by a SiLU activation function and adaptive average pooling. The first convolutional layer processes the input with 32 filters and a kernel size of 3x3, followed by a second convolutional layer with 64 filters and a third with 128 filters. Adaptive average pooling is applied after each convolutional layer to reduce spatial dimensions while preserving important features. Usually most CNN networks for classification have a classifier flollowed by Convolution layers based on Perceptron or multitilayer perceptron.

\subsection{ConvKAN with non-linear layer} \label{lab:convkan}
{ConvKAN with non-linear layer, illustrated in Fig. \ref{fig:convkan},  combines convolutional layers with the principles of KAN. It uses multiple ConvKAN layers with varying input and output channels, kernel sizes, and strides to extract hierarchical features from input images. Batch normalization is applied after each ConvKAN layer to stabilize training, and adaptive average pooling reduces spatial dimensions to a fixed size (1x1) before flattening. Then a fully connected layer is added with a SiLU activation function, which breaks the linearity, making the whole stack nonlinear. Basically, we can state that the output of the final ConvKAN layer is flattened into a one-dimensional vector. 
However, it is important to clarify the operations occurring within each convolutional block: after every convolution, a SiLU activation is applied, followed by a KAN layer that uses a learned B-spline activation function.

The key characteristic of the KAN layer is that it does not change the size or dimensions of the data—meaning it preserves the width, height, and number of channels of the feature maps. Instead, it transforms the values at each position in the feature maps using an adaptive nonlinear function, specifically a B-spline activation that is learned during training.

Basically, for each input element (pixel), the KAN layer applies a B-splines that is significantly more flexible and smoother than conventional activations such as ReLU, Sigmoid, or SiLU. So, the output shape remains unchanged, but the values are adaptively and “intelligently” transformed. The KAN layer thus enhances the quality of extracted features before they are passed to the classification layers.

\begin{figure}[h!]
\centering
\includegraphics[width=0.48\textwidth]{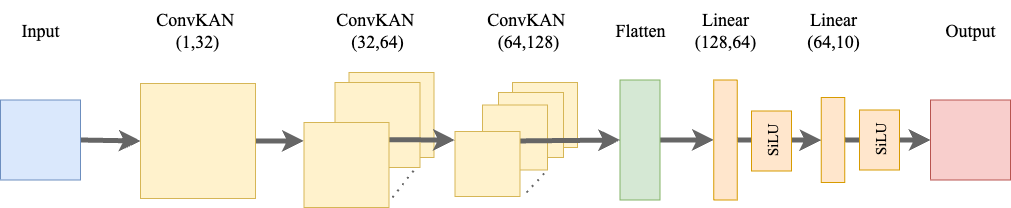}  
\caption{Architecture of ConvKAN with a non-linear structure}
\label{fig:convkan}
\end{figure}

\subsection{CapsNet}
CapsNet captures spatial hierarchies using capsules and dynamic routing. As shown in Fig. \ref{fig:capsnet}, it begins with a feature extractor to process input images. The PrimaryCaps layer transforms these features into capsules \cite{xi2017capsule}, applying the squash activation function to encode spatial information like pose and orientation.

\begin{figure}[h!]
\centering
\includegraphics[width=0.48\textwidth]{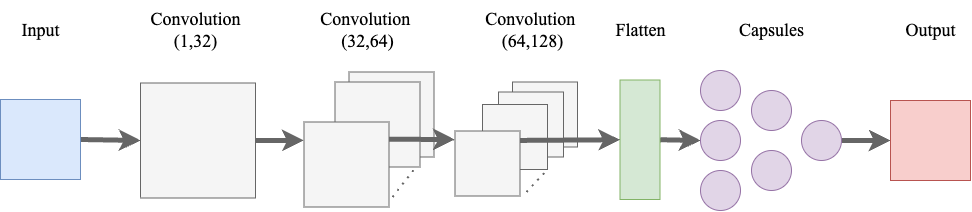}  
\caption{Architecture of Capsule Network}
\label{fig:capsnet}
\end{figure}

Capsules are refined through Routing Layers, which use iterative dynamic routing to model spatial relationships \cite{sabour2017dynamic}. The final Routing Layer produces output capsules representing class probabilities, with vector lengths indicating confidence \cite{dombetzki2018overview}. A dropout layer is included to prevent overfitting \cite{baldi2013understanding}.

\subsection{Capsule-ConvKAN}
Capsule-ConvKAN combines ConvKAN as the feature extractor with CapsNet, shown in Fig. \ref{fig:capsconvkan}. The model consists of multiple ConvKAN layers followed by batch normalization and adaptive average pooling. CapsNet captures spatial relationships using dynamic routing, with primary capsules converting features into capsules, and routing layers refining connections between capsules.

\begin{figure}[h!]
\centering
\includegraphics[width=0.48\textwidth]{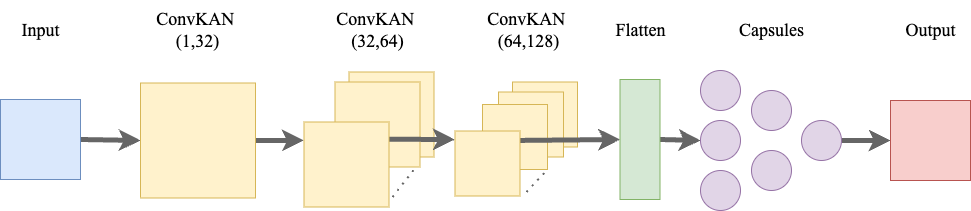}  
\caption{Capsule-ConvKAN}
\label{fig:capsconvkan}
\end{figure}

As described in Section \ref{lab:convkan}, the ConvKAN layer first processes the input feature maps through convolution followed by adaptive nonlinear transformations. The output of this final ConvKAN layer is then flattened into a one-dimensional vector, which is passed to the capsule classifier for further processing and classification.

\section{RESULTS}
The model’s performance is evaluated using accuracy, precision, F1 score, recall (sensitivity), specificity, AUC. These metrics provide comprehensive insights into classification errors and overall model effectiveness. In particular, within the healthcare domain, the inclusion of specificity, sensitivity, and AUC is crucial for accurately assessing the model’s diagnostic capabilities and ensuring reliable decision-making.

\subsection{Dataset Description}

The dataset, depicted in Fig.~\ref{fig:pcam}, used in this study consists of histopathological images of tissue samples stained with hematoxylin and eosin, a technique in medical imaging for highlighting cellular structures \cite{cao2021pcam}. 

\begin{figure}[!h]
\centering
\includegraphics[width=0.35\textwidth]{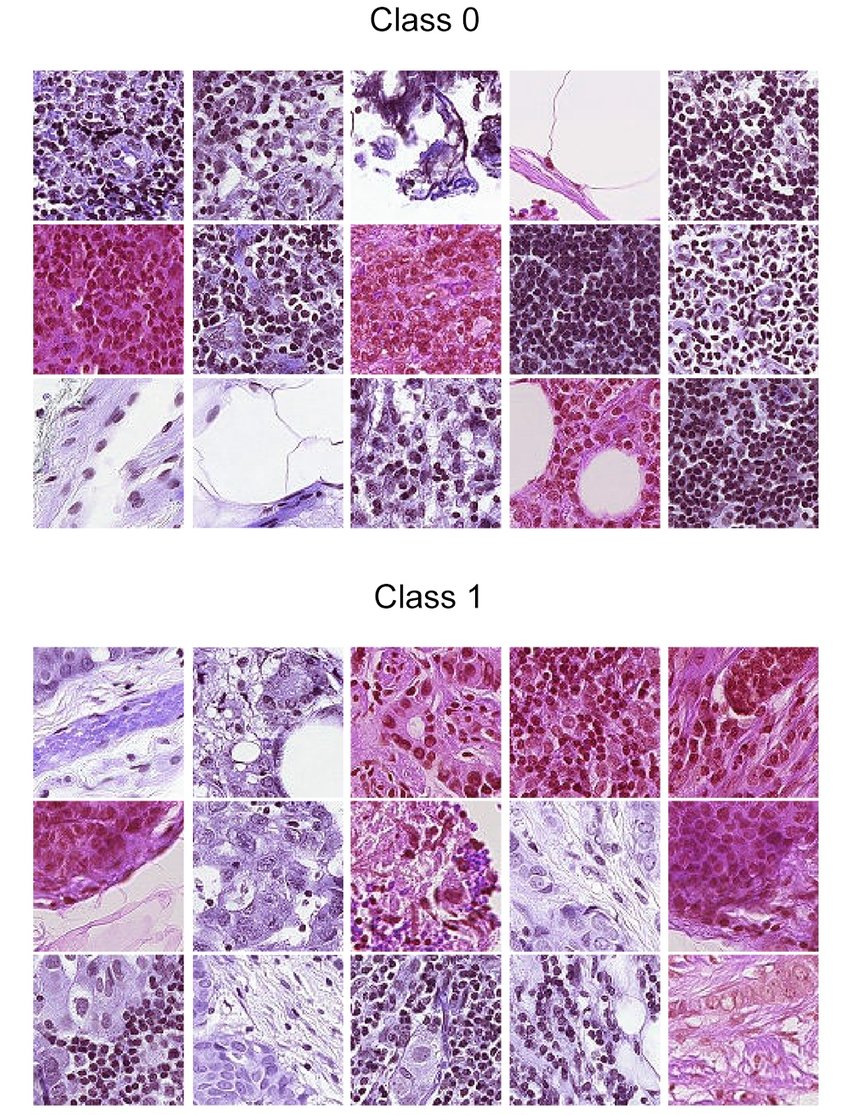}  
\caption{Example of dataset \cite{palatnik2019local}}
\label{fig:pcam}
\end{figure}

In this study, the dataset is categorized into two classes based on histological characteristics. The first, Class 0, includes tissue samples with normal or benign features, typically showing well-organized cellular structures, uniform distribution of nuclei, and minimal abnormalities. The second, Class 1, comprises malignant tissue samples, which often exhibit irregular cell morphology, an increased nuclear-to-cytoplasmic ratio, and densely clustered cells—hallmarks of cancerous histopathology.

For model training and evaluation, the dataset was split into training, validation, and test subsets. Specifically, 70\% of the data was allocated for training, 15\% for validation, and 15\% for testing. To ensure that both classes were proportionally represented across all subsets, a stratified splitting approach was employed. 

\begin{table}[h!]
\centering
\begin{tabular}{lcccc}
\hline
\textbf{Model} & \textbf{Accuracy} & \textbf{AUC} & \textbf{Specificity} & \textbf{Sensitivity} \\ \hline
CNN     & 85.60\% & 0.896 & 86.3\% & 85.55\% \\
ConvKAN & 89.53\% & 0.942 & 90.7\% & 87.21\% \\
CapsNet & 88.12\% & 0.927 & 89.4\% & 87.68\% \\
\textbf{Capsule-ConvKAN} & \textbf{91.21\%} & \textbf{0.951} & \textbf{92.2\%} & \textbf{89.56\%} \\ \hline
\end{tabular}
\caption{Comparison of model performance}
\label{tab:idc_results}
\end{table}

As shown in Table~\ref{tab:idc_results}, the Capsule-ConvKAN model achieves the highest accuracy (91.21\%), AUC (0.951), and specificity (92.2\%) compared to ConvKAN, CNN, and CapsNet.
In addition to these metrics, Capsule-ConvKAN also outperforms other models in precision (91.87\%), recall (89.56\%), and F1-score (90.41\%). For comparison, ConvKAN attains 89.53\% accuracy, 0.942 AUC, 90.7\% specificity, 88.39\% precision, 87.21\% recall, and 81.80\% F1-score; CNN achieves 85.60\% accuracy, 0.896 AUC, 86.3\% specificity, 80.73\% precision, 85.55\% recall, and 83.48\% F1-score; and CapsNet reaches 88.12\% accuracy, 0.927 AUC, 89.4\% specificity, 85.42\% precision, 87.68\% recall, and 86.51\% F1-score.
These results demonstrate that combining ConvKAN and capsule-based mechanisms yields more effective image classification performance on the IDC dataset.

\section{DISCUSSION}
The experimental results demonstrate that the Capsule-ConvKAN architecture outperforms the other compared models -- CNN, CapsNet, and ConvKAN -- achieving the highest accuracy on the histopathological dataset. This outcome validates the hypothesis that integrating the dynamic routing and spatial hierarchy modeling capabilities of capsule networks with the flexible, interpretable function approximation of ConvKAN creates a more powerful and robust architecture for image classification tasks.

The superior performance of Capsule-ConvKAN can be attributed to its ability to effectively capture spatial relationships and pose information through capsules, which is particularly beneficial for medical histopathological images where subtle structural differences indicate malignancy. Moreover, the underlying KAN framework allows the network to approximate complex functions more precisely using learnable B-spline activation functions, contributing to improved generalization.

In contrast, while ConvKAN alone leverages the Kolmogorov-Arnold representation to provide interpretability and function approximation flexibility, it lacks the capsule’s mechanism to model hierarchical part-whole relationships, resulting in lower performance. Traditional CNNs, despite their proven effectiveness in visual tasks, tend to lose spatial hierarchies due to pooling and convolution operations, which capsule networks address more explicitly.

The findings align with previous studies indicating that capsule networks enhance robustness to spatial transformations and deformations, critical in real-world applications such as medical image analysis where variations in tissue presentation occur frequently. Additionally, the use of learnable B-splines within KAN layers introduces interpretability and adaptability not commonly present in standard CNN architectures.

However, the increased complexity of Capsule-ConvKAN comes with higher computational costs and training times, which should be considered when deploying in resource-constrained environments. In this study, the computational power of the available PC limited the extent of hyperparameter tuning and experimentation with larger model configurations. Consequently, future work will be conducted on more powerful hardware to explore these aspects more comprehensively and potentially improve model performance.

Finally, the current study was conducted on a histopathological dataset; further validation on larger and more diverse medical image datasets would strengthen the generalizability of the results. Investigating the model’s explainability and interpretability in clinical decision-making contexts is also a promising direction.

\section{CONCLUSION}
The performance of Capsule-ConvKAN underscores the critical importance of integrating spatial awareness and hierarchical feature representation in image classification tasks. By combining the strengths of CapsNet and convolutional kernels, this hybrid architecture effectively captures complex patterns and relationships within histopathological images, leading to improved classification accuracy compared to traditional methods. These findings not only contribute valuable empirical evidence supporting the utility of hybrid deep learning models but also offer a strong foundation for further exploration and innovation in the field of medical image analysis.

Moreover, the successful application of Capsule-ConvKAN highlights the potential for such architectures to address challenges associated with subtle feature variations and spatial hierarchies inherent in complex image datasets. This advancement represents a meaningful step toward more robust, interpretable, and scalable image classification systems that can be adapted across various domains beyond histopathology.

Looking forward, future research will focus on extending the applicability of Capsule-ConvKAN to more diverse and complex datasets, including multi-modal and higher-resolution images. Efforts will also be directed toward optimizing the model’s scalability and computational efficiency to facilitate its integration into real-world clinical and diagnostic workflows. Additionally, exploring the interpretability of the learned representations within Capsule-ConvKAN could provide deeper insights into the decision-making processes of hybrid neural networks, further enhancing their trustworthiness and adoption in critical applications.

\bibliographystyle{unsrt}

\end{document}